\documentclass[twocolumn,PRL,showpacs,nofootinbib,amssymb]{revtex4-1}

\usepackage{amsmath,amsfonts}
\usepackage{dcolumn}
\usepackage{graphicx}
\usepackage{xcolor}
\usepackage{bm}
\usepackage{natbib}
\usepackage{calligra}
\usepackage[T1]{fontenc}
\usepackage{egothic}
\usepackage[T1]{fontenc}
\newfont{\rsfsten}{rsfs10 scaled 1200}
\newfont{\rsfsseven}{rsfs10 scaled 1200}
\newfont{\rsfsfive}{rsfs10 scaled 1200}
\usepackage{epsfig}
\usepackage{units}
\usepackage[utf8]{inputenc}

\newcommand{\mnras}{{Mon.~Not.~R.~Astron.~Soc.}}
\newcommand{\pasj}{{Publ.~Astron.~Soc.~Japan}}

\newcommand{\be}{\begin{equation}}
\newcommand{\ee}{\end{equation}}
\newcommand{\bea}{\begin{eqnarray}}
\newcommand{\eea}{\end{eqnarray}}

 \usepackage{ulem}

\def\lsim{\mathrel{\raise.3ex\hbox{$<$\kern-.75em\lower1ex\hbox{$\sim$}}}}
\def\gsim{\mathrel{\raise.3ex\hbox{$>$\kern-.75em\lower1ex\hbox{$\sim$}}}}

\newcommand{\class}{{\sc {class}}}

\definecolor{darkred}{RGB}{175,0,0}

\begin{document}

\title{
Effects of primordial black holes quantum gravity decay on galaxy clustering
}

\author{Alvise Raccanelli
$^\star\let\thefootnote\relax\footnote{$^\star$Marie Sk\l{}odowska-Curie fellow}$$^{1,2}$\thanks{miscellaneous text}, Francesca Vidotto$^3$, Licia Verde$^{1,4}$}
\affiliation{
$^1$ Institut de Ci\`encies del Cosmos (ICCUB), Universitat de Barcelona (IEEC-UB), Mart\'{i} Franqu\`es 1, E08028 Barcelona, Spain \\
$^2$ Department of Physics \& Astronomy, The Johns Hopkins University, Baltimore, Maryland, 21218, USA \\
$^3$ Institute for Mathematics, Astrophysics and Particle Physics (IMAPP), Radboud University,  P.O. Box 9010, 6500 GL Nijmegen, The Netherlands \\
$^4$ ICREA, Pg. Lluis Companys 23, 08010 Barcelona, Spain
}
\date{\today}

\begin{abstract}
It has been recently suggested that small mass black holes (BHs) may become unstable due to quantum-gravitational effects and eventually decay, producing radiation, on a timescale shorter than the Hawking evaporation time.
We argue that the existence of a population of low-mass Primordial Black Holes (PBHs) acting as a fraction of the Universe dark matter component can be used to test proposed models of quantum decay of BHs 
via their effect on galaxy number counts.
We study what constraints future galaxy clustering measurements can set on quantum-gravity parameters governing the BH lifetime and PBH abundance.
In case of no detection of such effects, this would rule out either the existence of a non-negligible number of small PBHs, or the BH quantum decay scenario (or both).
In case of independent observations of PBHs, the observables discussed here could be used to study the quantum effects that modify the final fate of BHs.
\end{abstract}


\maketitle

\section{Introduction}
\label{sec:intro}
\vspace{-0.15cm}
Understanding the connection between General Relativity and quantum mechanics is one of the deepest open issues in physics.
Our knowledge of the quantum properties of the gravitational field is still not satisfactory, and the need for empirical tests is especially urgent. Black holes (BHs) have always been a privileged framework to probe ideas in quantum gravity (QG). On the one hand, there are general qualitative arguments, shared by different approaches, that can give hints for possible phenomenological effects; on the other hand, new developments in the theory may provide concrete calculations. 

Black holes are challenging our understanding of their formation mechanism, 
as they are being discovered in a wide mass range. The recent observations by the LIGO instrument~\cite{Abbott:2016blz, Abbott2, Abbott3}, while confirming the presence of gravitational waves, have opened new questions about the origin of black holes of such masses.
First studied in~\cite{Zeldovich:1967, Hawking:1971, Carr:1974}, Primordial Black Holes (PBHs) are BHs formed at very early times not as endpoint of stellar evolution, but as a consequence of different possible mechanisms.
Therefore, a speculation comes to mind: has LIGO detected primordial black holes~\cite{Bird:2016dcv}?

The idea that DM could be constituted by PBHs dates back to the Seventies~\cite{Chapline:1975cr}. Since then, observational consequences and constraints have been considered for different PBH mass ranges and for different observables (see e.g.~\cite{Carr:2016, Green:2016xgy, Kuhnel:2017, Carr:2017}). Ways to test this scenario were proposed in e.g.~\cite{Raccanelli:2016BH, Cholis:2016, Munoz:2016, Raccanelli:2016GW, Kovetz:2016BHMF, Kovetz:2017}.

The sub-solar mass part of the PBH parameter space is mostly constrained by microlensing, and on even smaller masses, constraints come from observed limits on high-energy cosmic rays that would be produced at the end of the Hawking evaporation~\cite{Barrau:1999sk}.
In this paper we consider a different possibility: Hawking evaporation may not be the most relevant phenomenon to constrain small mass PBHs, if QG physics modifies the final fate of black holes before evaporation becomes significant.

Quantum fluctuations of the metric can become important outside the black hole horizon, as first emphasized in~\cite{Dvali:2011aa,Dvali:2017eba,Giddings:2017xh}. Here we consider the possibility that quantum fluctuations could yield the disruption of the horizon and that eventually  all the black-hole mass  is converted into radiation (and possibly gravitational waves) on a time scale shorter that the standard Hawking evaporation time (for one possible speculative scenario see e.g.~\cite{Rovelli:2014cta,Rovelli:2017mw}). 

The current interest in the PBH as DM model, the recent developments concerning the possibility of quantum fluctuations of the metric near the BH horizon, and the advent of extremely precise large-scale galaxy surveys, make it timely to investigate the possibility of extending tests of such fundamental models with forthcoming experiments.

In this work we argue that the existence of a population of PBHs of small mass that could make up a non-negligible fraction of the dark matter would imply that QG BH decay could affect galaxy clustering measurements.

This paper is organized as follows. In Section~\ref{sec:QBH} we review the theoretical motivations for considering the decay of BHs triggered by QG effects.
In Section~\ref{sec:lss} we discuss how, in the model where PBHs make up DM, the variation of the effective DM energy density due to such a decay can be measured by considering the modifications of galaxy clustering and cosmic magnification due to DM gravitational lensing.
Our results are presented in Section~\ref{sec:results}, and discussed in Section~\ref{sec:conclusions}.

\section{Quantum effects in Black Holes {and decay}}
\label{sec:QBH}
In classical General Relativity, black holes are mathematically characterised by an event horizon that, once  formed, never disappears; however, at relevant scales, this picture should be corrected by quantum mechanics. Hawking evaporation provides a mechanism that shrinks the horizon, now a 
{\it dynamical} horizon, 
on a timescale proportional to the cube of the mass of the black hole $M_{\rm BH}^3$ (in natural units). For astrophysical black holes, this long timescale implies a practical observational irrelevance. For PBHs, which could be formed in a very wide mass range, it produces possibly relevant effects only for PBHs of mass below $10^{12}$ \,Kg~\cite{Carr:1976zz}: this is the mass that would correspond to PBHs evaporating today.

Hawking evaporation is a well known phenomenon that modifies the classical prediction, but other quantum phenomena can in principle appear on a shorter timescale. Generic classical systems have a characteristic timescale after which their evolution departs from the classical one and quantum effects appear. 
This is also true in QG. 
For a BH, this time could in principle be shorter than the Hawking evaporation time, because the fact that the approximation given by Quantum Field Theory on a fixed (curved) background, plus backreaction, is likely to already break down when the mass of the hole enters the quantum regime;
this is a reasonable physical interpretation of the `firewall' theorem~\cite{Almheiri:2012rt}. Specifically, significant quantum fluctuations of the metric can appear in the region outside the horizon after a time as short as $M_{\rm BH}^2$, and this would have important consequences for BH phenomenology.

Quantum fluctuations of the metric are a typical manifestation of QG. 
On the horizon of a macroscopic BH, the curvature is small and proportional to $M_{\rm BH}^{-2}$. The inverse of this quantity defines a timescale that corresponds to the time needed for quantum effects to manifest themselves outside of the horizon~\cite{Haggard:2014fv, Haggard:2016ibp}. This may yield dramatic changes to the horizon, possibly making the BH decay 
into a different state of the geometry without a BH horizon, so that radiation can be released.

For example, instabilities due to QG yielding BH disappearance in a timescale shorter than the evaporation time have been explored in different approaches to QG~\cite{Gregory:1993vy, Casadio:2001dc, Casadio:2000py, Kol:2004pn, Emparan:2002jp}. In particular, an explicit calculation of the BH lifetime has become available in the context of Loop Quantum Gravity (LQG)~\cite{Christodoulou:2016ve}.

 The methodology and observables we suggest in the next Section are valid for any theoretical framework predicting 
 BH decay, or any phenomenological models for 
 QG effects with a similar behaviour.
For concreteness, let us consider here an example scenario, that we take as a test case, to compute our results.
In this scenario, spacetime discreteness in LQG implies the absence of curvature singularities~\cite{Rovelli:2013osa, Ashtekar:2005qt, Corichi:2015xia}. QG effects prevent the formation of the singularity and trigger a new expanding phase~\cite{Rovelli:2014cta,Haggard:2014fv},  
which eventually leads to the decay of the BH  
and the release of radiation~\cite{Barrau:2014qd,Barrau:2014bv,Barrau:2015uca,Barrau:2016fcg,Vidotto:2016jqx}.

Computing the exact value of the BH lifetime provides a challenge for different QG approaches, and may provide a test for them. For the sake of phenomenology, a window of allowed lifetime has been considered: this ranges roughly from the minimal time $M_{\rm BH}^2$, a possibly favourite value, to $M_{\rm BH}^3$. We parameterize the BH lifetime $\tau$ as:
\begin{equation}
\label{eq:alpha}
\tau = \left(\frac{M_{\rm BH}}{m_{\rm pl}}\right)^{\alpha} t_{\rm pl} \, ,
\end{equation}
where $t_{\rm pl}$ and $m_{\rm pl}$ are the Planck time and mass, and $M_{\rm BH}$ is the BH mass; throughout this paper we will assume $\alpha=2$ as a fiducial value. 
Here we use Equation~\eqref{eq:alpha} to constrain the masses of decaying PHBs; however, in presence of an independent detection of PBHs, our results can be reinterpreted to constrain different values of the quantum parameter $\alpha$.

\section{Methodology}
\label{sec:lss}
We investigate one of the observational consequences of BH decay, as described by Equation~\eqref{eq:alpha}, regardless of the physical mechanism driving it.
If small mass ($\lesssim 10^{-15} M_\odot$) PBHs exist in non-negligible number (so that they might constitute at least part of the DM) and they decay, they will effectively convert part of the matter energy-density into radiation\footnote{The details about the conversion have been investigated in~\cite{Barrau:2014qd,Barrau:2014bv,Barrau:2015uca,Barrau:2016fcg,Vidotto:2016jqx}
, and are not important for the present analysis; more observational consequences will be investigated in a follow up paper in preparation~\cite{Bellomo:prep}.} yielding a net decrease of an effective dark matter density parameter $\Omega_{\rm DM}^{\rm eff}$, mimicking in practice a decaying (or annihilating) DM component (for which some cosmological consequences have been studied in e.g.~\cite{Aoyama:2014, Poulin:2016}).
This leaves a characteristic imprint on several cosmological observables, that we now introduce.

\subsection{Galaxy clustering}
As DM drives the growth of cosmological perturbations, an effective decaying DM component will leave a signature on the clustering of galaxies. When making observations on our past light-cone, we observe angular positions and redshifts which are perturbed by peculiar velocities~\cite{Kaiser:1987qv}, but also gravitational lensing and potentials (see e.g.~\cite{Raccanelli:2016GR}).
We can then define the {\it observed} galaxy over-density $\Delta_{\rm obs}({\bf n},z)$ at fixed observed redshift and into a given observed direction ${\bf n}$. 

We can therefore express the total {\it observed} over-density as:
\begin{equation}
\label{eq:deltas}
\Delta_{\rm obs} = \Delta_{\delta} + \Delta_{\rm rsd} + \Delta_{\rm v} + \Delta_{\rm \kappa} + \Delta_{\rm pot} + \Delta_{\mathfrak{t}} \, ,
\end{equation}
where $\delta$ refers to the overdensity in the comoving gauge, $rsd$ and $v$ denote  Doppler effects and peculiar velocity ~\cite{Raccanelli:Doppler},
$\kappa$ the lensing convergence and $pot$ incorporates local and non local terms depending on Bardeen potentials and their temporal derivatives; $\mathfrak{t}$ includes tensor perturbation effects. For simplicity, we omitted the redshift and direction dependencies $({\bf n}, z)$.

Galaxy clustering measurements have been used to measure cosmological parameters for a variety of cosmological models (see e.g.~\cite{Samushia:2012, Raccanelli:2013, Ross:2013, Samushia:2014, Sanchez:2014, Aubourg:2015, Alam:2016}), and it is the focus of many future large scale galaxy surveys (see e.g.~\cite{Raccanelli:radio, PFS, SKA:RSD, Euclid, DESI}). Therefore it is interesting to see that it can also be used to test different QG models and potentially extend constraints on PBHs as DM.

The most relevant statistical quantity usually measured is the 2-point function; its spherical harmonic counterpart,  the angular power spectrum, that correlates two probes $X$ and $Y$ is:
\begin{equation}
C_{\ell}^{XY}(z_i,z_j)=\left< a^X_{\ell m}\!(z_i) \ a^{Y\,^*}_{\ell m}\!(z_j) \right> \;,
\end{equation}
where the star denotes complex conjugation, and the spherical harmonics coefficients  are defined by $X=\sum a_{\ell m}{\cal Y}_{\ell m}(\bf n)$, with $ {\cal Y}_{\ell m}$ denoting the spherical harmonics functions, ${\bf n}$ the direction on the sky and $z_i$ is the redshift. This can be calculated from the underlying matter power spectrum by using:
\begin{align}
\label{eq:ClXY}
C_{\ell}^{XY}(z_i,z_j) = \int \frac{4\pi dk}{k} \ \Delta^{\!2}\!(k) \ W_{\ell}^X\!(k, z_i)\  W_{\ell}^Y\!(k, z_j) \, ,
\end{align}
where $W_{\ell}^{\{X,Y\}}$ are the source distribution kernels 
for the different observables (i.e. galaxies in different redshift bins) and $\Delta^2(k)$ 
is the dimensionless matter power spectrum today.

The kernel for the galaxy clustering can be written as 
(see e.g.~\cite{Nolta:2004}):
\begin{equation}
\label{eq:flg}
W_{\ell}^X\!(k) = \int \frac{d N_X(z)}{dz} \ D(z) \ b_X\!(z) \  j_{\ell}[k\chi(z)] \ dz \, ,
\end{equation}
where $d N_X(z)/dz$ is the objects redshift distribution, $D(z)$ the growth rate of structures,
$b_X(z)$ is the bias that relates the observed overdensity to the underlying matter distribution (see~\cite{biasreview} for a recent review);
$j_{\ell}(x)$ is the spherical Bessel function of order $\ell$, and $\chi(z)$ is the comoving distance. \\

\subsection{Cosmic Magnification}
Gravitational lensing causes the deflection of light rays by the matter distribution along the line of sight, causing two competing effects: on one hand, a size magnification of sources behind a lens, on the other hand lensing causes the stretching of the observed field of view.
Therefore, for magnitude (or flux) limited galaxy surveys, sources that are just below the threshold for detection will be magnified and become detectable, so that the observed number density of sources increases. At the same time, the stretching of the field of view leads to a decrease of the observed number density. The combined effect can be written as:
\begin{equation}
\label{eq:cosmag}
n^{\rm obs}(z) = n_g(z) [1 + (5s-2) \kappa] \, ,
\end{equation}
where $n^{\rm obs}$ and $n_g$ are the observed and intrinsic number of sources, respectively, $s$ is the magnification bias and $\kappa$ is the convergence.
The net modification of the observed number density is called magnification bias $s$:
\begin{eqnarray}
s = \frac{d {\,\rm log} N_{|_{< M}}}{d M}\bigg|_{M_{\rm lim}} \, ,
\end{eqnarray}
where $M_{\rm lim}$ is the magnitude (or flux) limit of the survey and $N_{|_{< M}}$ is the number count for galaxies brighter than a magnitude (or flux) $M$.

Cosmic magnification has been suggested as a probe for cosmology~\cite{Matsubara:2000} and has been subsequently studied in a variety of works (see e.g.~\cite{Hui:2007, Loverde:2008}); as part of the contributions to the {\it observed} large-scale galaxy correlation, it has been analyzed in e.g.~\cite{Raccanelli:radial, Montanari:2015, Raccanelli:2016GR}.
From Equation~\eqref{eq:ClXY},  it is clear that cosmic magnification could be detected by cross-correlating galaxies in two disjoint redshift bins (see e.g.~\cite{Kaiser:1998, Scranton:2005}).

\subsection{Galaxy surveys}
\begin{figure}[h!]
\begin{centering}
\includegraphics[width=\columnwidth]{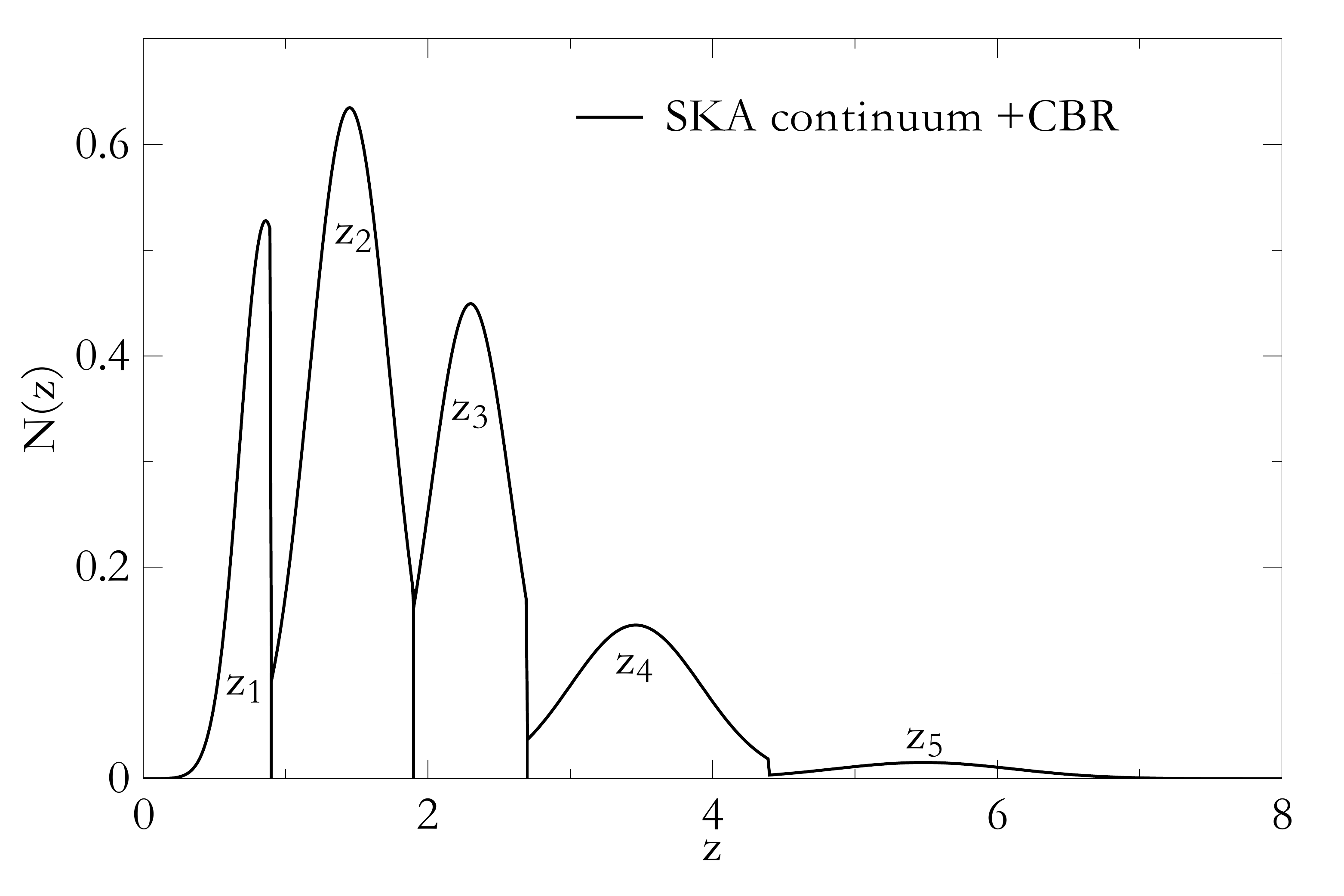}
\end{centering}
\caption{Adopted redshift distribution (normalized to 1) for the continuum SKA galaxy survey with  clustering based redshift (CBR). Each of the  peaks correspond to a redshift bin as indicated by the labels.
}
\label{fig:Nz}
\end{figure}

We model our galaxy survey after the Square Kilometre Array (SKA)\footnote{\url{https://skatelescope.org}}, which is an international multi-purpose next-generation radio interferometer, that will be built in the Southern Hemisphere in Africa and in Australia, with a total collecting area of about 1 km$^2$.
Among many types of observations delivered by such instruments, we focus here on surveys that will detect individual galaxies in the radio continuum~\cite{SKA:continuum}; we assume that the survey will cover $30,000\;\text{deg}^2$, and we compute results for a flux limit of $1\mu Jy$.
Although radio continuum surveys do not have in principle redshift information, some techniques have been proposed to allow the possibility to divide the galaxy catalog into tomographic redshift bins; here we follow the clustering-based redshift (CBR) information approach proposed in~\cite{Menard:2013}, and studied for some cosmological applications (including some predictions for the SKA), in~\cite{Kovetz:2016}.
In Figure~\ref{fig:Nz} we show the (normalized) redshift distribution for the SKA radio continuum survey we use for this paper.

We compute the observational consequences of BH decay by using a modified version of the \class{}\footnote{{\url{http://class-code.net/}}} code, and we investigate the effects on galaxy angular power spectra.
Given the specifications of the proposed future surveys, we forecast the measurements' precision using the Fisher matrix approach~\citep{Fisher:1935, Tegmark:1997}; the Fisher matrix is:
\begin{equation}
\label{eq:Fisher}
F_{\alpha\beta} = \sum_{\ell} \ 
\frac{\partial C_\ell}{\partial \vartheta_\alpha} \
\frac{\partial C_\ell}{\partial \vartheta_\beta}\ 
{\sigma_{C_\ell}^{-2}} \, , 
\end{equation}
where $\vartheta_{\{\alpha, \beta\}}$ are the parameters one wants to measure and the derivatives of the power spectra $C_\ell$ are evaluated at fiducial values $\bar \vartheta_{\alpha}$.
In this work we assume Planck+BOSS priors over a $\Lambda$CDM flat- cosmological constant cosmology, and we marginalize over all cosmological parameters apart from the BH lifetime, which is then translated into a $M_{\rm BH}$ via Equation~\eqref{eq:alpha}.
The quantities $\sigma_{C_\ell}$ are measurement errors in the power spectra, which we compute as (see e.g.~\cite{Raccanelli:2016GR}):
\begin{equation}
\label{eq:clgt}
\sigma^2_{C_{\ell \, [(ij), (pq)]}} = \frac{\tilde{C}_{\ell}^{ (ip)} \tilde{C}_{\ell}^{ (jq)} + \tilde{C}_{\ell}^{ (iq)} \tilde{C}_{\ell}^{ (jp)}}{(2\ell+1)f_{\rm sky}} \, ,
\end{equation}
where $f_{\rm sky}$ is the fraction of the sky surveyed and $\tilde{C}_{\ell} $ is the observed power spectrum, including the shot noise:
\begin{equation}
\label{eq:err-clgt}
\tilde{C}_{\ell}  = C_{\ell}^{ij} + \frac{\delta_{ij}}{dN(z_i)/d\Omega} \,,
\end{equation}
with $dN(z_i)/d\Omega$ being the average number of sources per steradian within the bin $z_i$.
We sum over the matrix indices $(ij)$ with $i\leq j$ and $(pq)$ with
$p \leq q$ which run from 1 to the number of bins; in our case we use the 5 bins of Figure~\ref{fig:Nz}.

\vspace{-0.25cm}

\section{Results}
\label{sec:results}
\begin{figure}[htb!]
\includegraphics[width=\columnwidth]{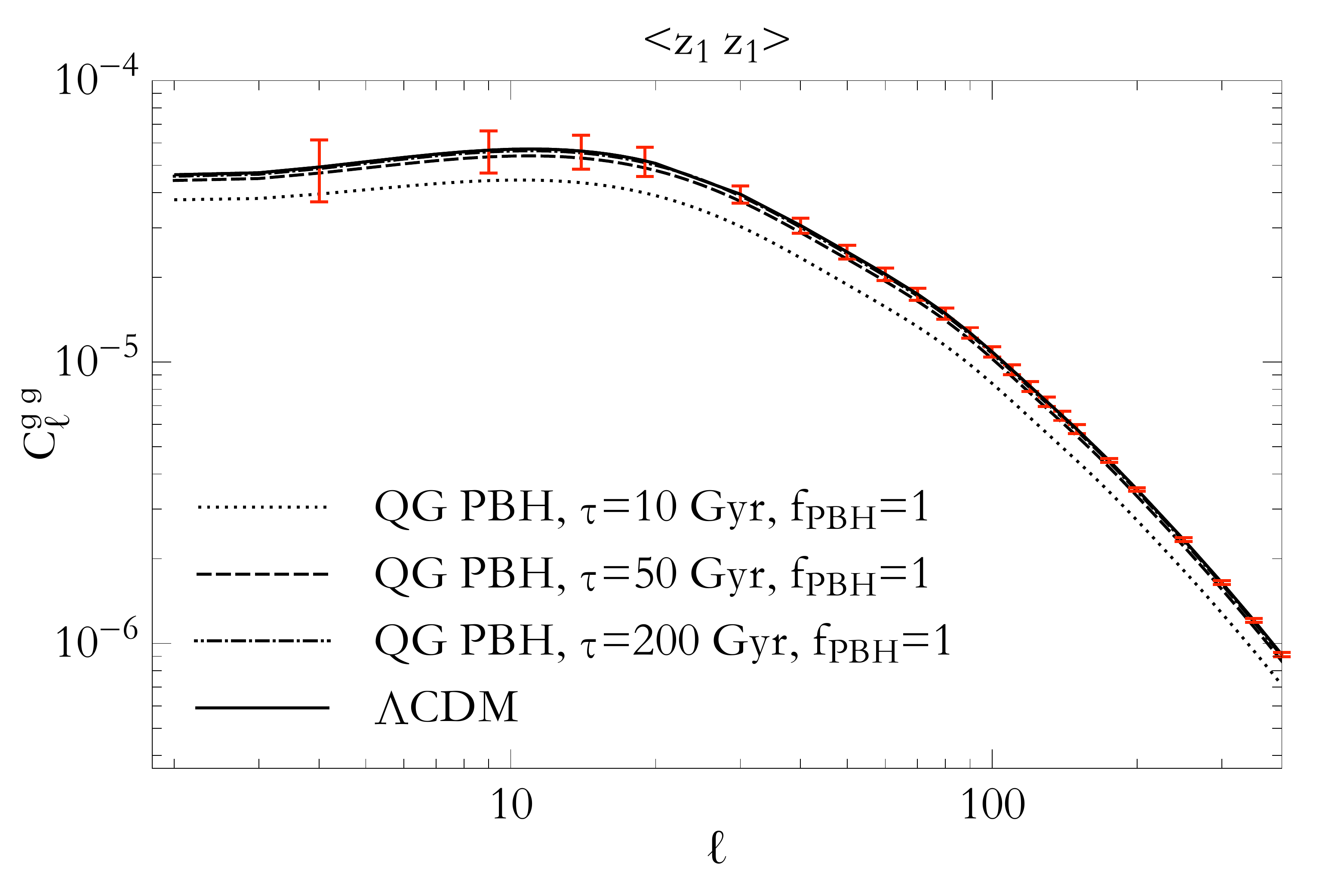}
\caption{Angular power spectrum for the SKA continuum 1$\mu$Jy, for the auto-correlation of galaxy clustering of the first z-bin. The solid line (with error bars, shown for clarity in bandpowers) shows the prediction for $\Lambda$CDM, while dotted, dashed and dot-dashed lines show the predictions for different values of PBH lifetimes as in legend; $f_{\rm PBH}$ is the fraction of DM in PBHs.
}
\label{fig:Clgg}
\end{figure}

\vspace{-0.15cm}
In Figure~\ref{fig:Clgg} we show the effect of PBH decay on the auto-correlation of the first redshift bin of the SKA continuum 1$\mu$Jy distribution of Figure~\ref{fig:Nz}; in this case we include the first three terms of the right-hand-side of Equation~\eqref{eq:deltas}, i.e. intrinsic clustering, RSD and velocity terms.
We plot the $\Lambda$CDM (solid line) and the PBH tunneling predictions for a series of BH lifetimes; every lifetime corresponds to a BH mass, as for Equation~\eqref{eq:alpha}. For illustrative purposes error bars are shown for bandpower measurements.
Here $f_{\rm PBH}$ is the fraction of DM in PBHs and, we compute results for monochromatic PBH masses.

\begin{figure}[h!]
\begin{centering}
\includegraphics[width=\columnwidth]{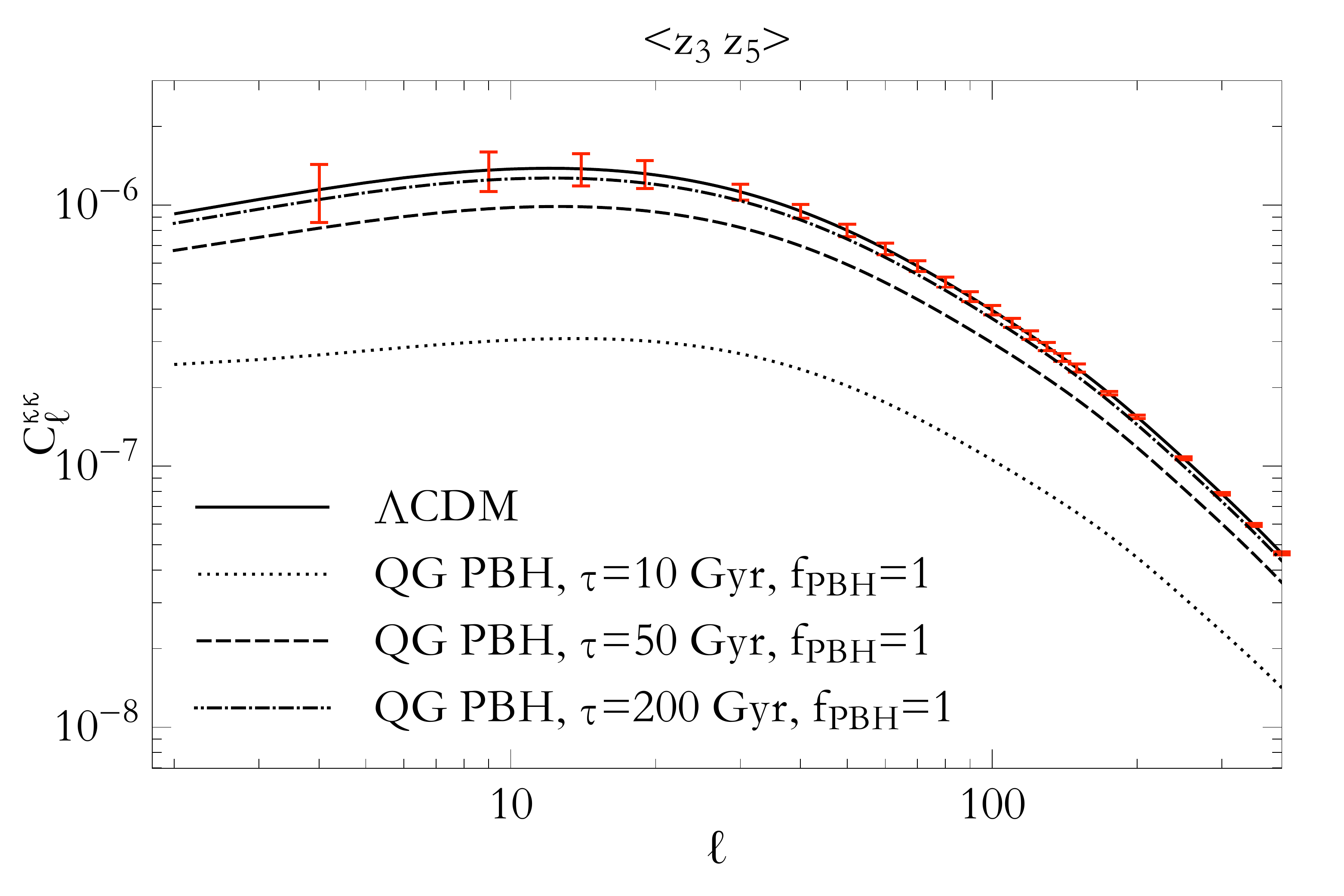}
\end{centering}
\caption{Angular power spectra for the radial cross-correlation of lensing convergence between the third and fifth bins. Legend as in Figure~\ref{fig:Clgg}.
}
\label{fig:Clkk}
\end{figure}

In Figure~\ref{fig:Clkk} we plot the radial cross-correlation of the third with the fifth bins of the $\kappa$ lensing contribution of Equation~\eqref{eq:deltas}.
The effect on the lensing term is in general larger than in the clustering case because the cross-correlation captures not only a change in the amplitude of clustering and the growth of structures at a specific time, but also the combination of geometry and redshift evolution.
In both cases we show, for clarity, only cases in which PBHs make up all the DM, while final constraints will be computed for any value of $f_{\rm PBH}$.

\begin{figure}[htb!]
\begin{centering}
\includegraphics[width=\columnwidth]{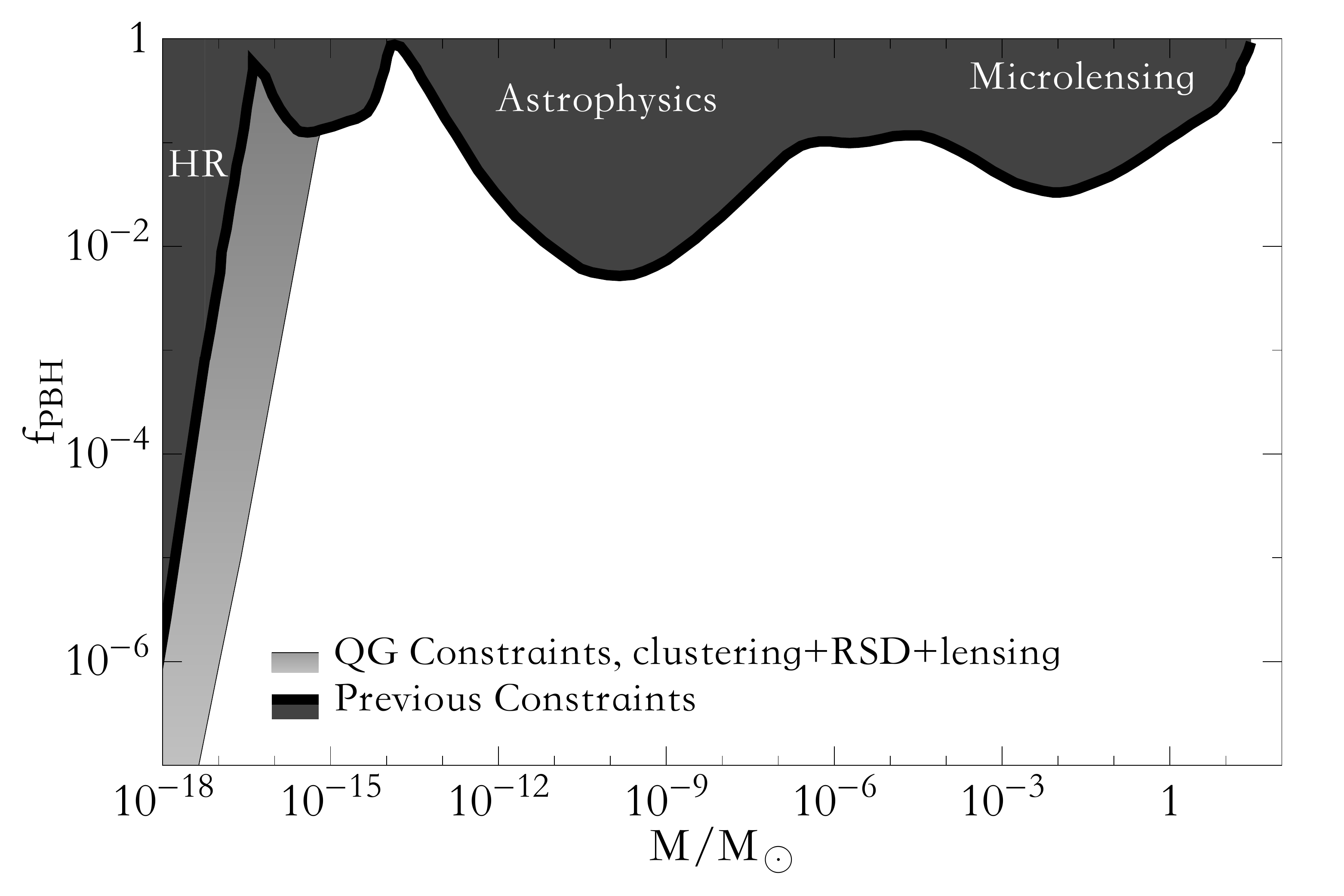}
\end{centering}
\caption{Predicted constraints (95\% confidence level) on the allowed mass and fraction of DM in PBHs from QG effects on galaxy clustering and lensing (light gray). Current constraints (dark gray) are  from~\cite{Carr:2017}.
}
\label{fig:constraints}
\end{figure}

We then performed a Fisher matrix analysis of the combination of all the auto- and cross- correlations to produce the total constraints on the allowed mass for PBHs as DM. We checked that different combination of observables or experiment configurations do not considerably affect our results; a careful comparison of results for different galaxy surveys is left for future works.

Assuming that the null hypothesis is that PBHs don't decay, we can compute constraints on the decay rate of the dark matter component, which is inversely proportional to the BH lifetime, which determines a mass (given a choice of $\alpha$) via Equation~\eqref{eq:alpha}.

In Figure~\ref{fig:constraints} we present the main result of the paper, the predicted constraints on the allowed mass of PBHs as DM and their abundance in presence of these decaying phenomena.
The thick black line shows a combination of current constraints (from~\cite{Carr:2017}), where $HR$ indicates constraints from Hawking radiation gamma-ray production; the area above the line represents the excluded fraction of PBHs as DM.
The light grey region represents the area that would be excluded in case of no detection of DM PBH decay, according to the assumptions made in this paper.
As one can see, once we take into account QG effects, the constraints from BH conversion into energy are considerably stronger than the ones due to HR and would reduce the allowed parameter space, in the small mass end, by more than one order of magnitude.
We can also note that our proposed methodology opens up the exciting possibility to test QG effects in BHs with forthcoming LSS data.

A few caveats need to be considered here. Not only we assume that PBHs are at least a fraction of the DM and that they decay,
we also assume a monochromatic mass distribution for the PBH and a value for $\alpha$.
We expect constraints not to  change drastically when varying  the assumptions on the value of $\alpha$ or the PBH mass distribution; our findings will be modified quantitatively but not qualitatively, and in this work we present a first indication that we can use galaxy clustering to constrain the PBH as DM model when considering QG effects.

\vspace{-0.25cm}
\section{Conclusions}
\label{sec:conclusions}
\vspace{-0.15cm}
We have pointed out the possibility to study QG parameters from measurements in the late universe. We have considered the (speculative) possibility that QG phenomena affect the behaviour of black holes yielding their decay, in a time proportional to a power of their mass and possibly shorter than the Hawking evaporation time. Therefore, if PBHs contribute to a fraction of the DM, there is an effect on the time evolution of the mass-energy density, which can be detected by galaxy number count measurements from large-scale galaxy surveys.

This effect is linked to BH lifetime, which depends on the decay mechanism as a function of the mass, here encoded by the effective parameter $\alpha$.
Therefore, it would represent both an additional observational constraint on the allowed mass of PBHs to be the DM, and a first test for QG parameters using galaxy clustering measurements.

We forecast what constraints could be set by using angular power spectra measurements from forthcoming galaxy surveys; our results indicate that then the excluded mass range for PBHs as DM could be considerably larger (in the small-mass part of the parameter space).

In the case of a detection of a non-negligible fraction of the DM to be comprised of PBHs (from different observables), limits from this type of analyses could provide a measurement of the QG lifetime-mass scaling. A more detailed study of this possibility (and the influence of relaxing the above assumptions) is currently undergoing.

To summarize, we argue that in a speculative scenario where PBHs contribute to the DM and QG effects lead to BH decay, then we will expect to detect a modification of galaxy number counts, or rule out PBHs as DM in a larger part of the parameter space compared to the HR $\gamma$-rays limits.
Conversely, a non-detection of this signal would signify that either PBHs in that range of masses are not the DM, or constrain QG effects (or both).

Additional observational constraints for the QG BH decay could come from energy injection on the CMB and gamma- and cosmic- ray signatures; a detailed investigation of these observables and a combined constraints analysis, including the fact that the PBHs can have an extended mass function, will be presented in a follow-up paper in preparation~\cite{Bellomo:prep}.

\bigskip
\noindent
{\bf Acknowledgements:}\\
The authors would like to thank Roberto Emparan and Mairi Sakellariadou for useful discussions. Aur\'elien Barrau, Nicola Bellomo, Jos\'e Luis Bernal, Ilias Cholis, Cesar Gomez, Daniel Holz and in particular Carlo Rovelli for inputs at the beginning of the project. We are grateful to Michael Turner for constructing feedback and for noticing a mistake in an earlier version of the draft. \\
AR has received funding from the People Programme (Marie Curie Actions) of the European Union H2020 Programme under REA grant agreement number 706896 (COSMOFLAGS). Funding for this work was partially provided by the Spanish MINECO under MDM-2014-0369 of ICCUB (Unidad de Excelencia ``Maria de Maeztu") and the Templeton Foundation. \\
FV's work at Radboud University is financed by the Innovational Research Incentives Scheme Veni of the Netherlands Organisation for Scientific Research (NWO). \\
LV acknowledges support of AYA2014-58747-P AEI/FEDER UE and of H2020 ERC 725327 BePreSysE project and warmly thanks the Radcliffe Institute for Advanced Study at Harvard University for hospitality.


\end{document}